\documentclass[sigconf]{acmart}
\AtBeginDocument{%
  \providecommand\BibTeX{{%
    \normalfont B\kern-0.5em{\scshape i\kern-0.25em b}\kern-0.8em\TeX}}}

\copyrightyear{2022}
\acmYear{2022}
\setcopyright{acmlicensed}\acmConference[UIST '22]{The 35th Annual ACM Symposium on User Interface Software and Technology}{October 29-November 2, 2022}{Bend, OR, USA}
\acmBooktitle{The 35th Annual ACM Symposium on User Interface Software and Technology (To Appear in UIST '22), October 29-November 2, 2022, Bend, OR, USA}
\acmPrice{15.00}
\acmDOI{XXXXXXX.XXXXXXX}
\acmISBN{978-1-4503-XXXX-X/18/06}

\usepackage{multirow}
\usepackage{fancyvrb}

\begin{document}

\title[Social Simulacra]{Social Simulacra: Creating Populated Prototypes \\ for Social Computing Systems}

\author{Joon Sung Park}
\affiliation{%
  \institution{Stanford University}
  \city{Stanford}
  \country{USA}}
\email{joonspk@stanford.edu}

\author{Lindsay Popowski}
\affiliation{%
  \institution{Stanford University}
  \city{Stanford}
  \country{USA}}
\email{popowski@stanford.edu}

\author{Carrie J. Cai}
\affiliation{%
  \institution{Google Research}
  \city{Mountain View, CA}
  \country{USA}}
\email{cjcai@google.com}

\author{Meredith Ringel Morris}
\affiliation{%
  \institution{Google Research}
  \city{Seattle, WA}
  \country{USA}}
\email{merrie@google.com}

\author{Percy Liang}
\affiliation{%
  \institution{Stanford University}
  \city{Stanford}
  \country{USA}}
\email{pliang@cs.stanford.edu}

\author{Michael S. Bernstein}
\affiliation{%
  \institution{Stanford University}
  \city{Stanford}
  \country{USA}}
\email{msb@cs.stanford.edu}

\renewcommand{\shortauthors}{J.S. Park, L. Popowski, C.J. Cai, M. Morris, P. Liang, M.S. Bernstein}

\newcommand{\gentxt}[1]{{\color{darkgray}\small\fontfamily{phv}\selectfont
#1}}

\newcommand{\msb}[1]{{\color{blue}{MSB: #1}}}
\newcommand{\lp}[1]{{\color{magenta}{Lindsay: #1}}}
\newcommand{\mlg}[1]{{\color{red}{Mitchell: #1}}}

\newcommand{\joon}[1]{{\color{purple}{#1}}}

\newcommand{\still}[1]{{\color{orange}{#1}}}

\newcommand{\mirev}[1]{{\color{black}{#1}}}

\begin{abstract}
    Social computing prototypes probe the social behaviors that may arise in an envisioned system design. This prototyping practice is currently limited to recruiting small groups of people. Unfortunately, many challenges do not arise until a system is populated at a larger scale. Can a designer understand how a social system might behave when populated, and make adjustments to the design before the system falls prey to such challenges? We introduce \textit{social simulacra}, a prototyping technique that generates a breadth of realistic social interactions that may emerge when a social computing system is populated. Social simulacra take as input the designer's description of a community’s design---goal, rules, and member personas---and produce as output an instance of that design with simulated behavior, including posts, replies, and anti-social behaviors. We demonstrate that social simulacra shift the behaviors that they generate appropriately in response to design changes, and that they enable exploration of ``what if?'' scenarios where community members or moderators intervene. To power social simulacra, we contribute techniques for prompting a large language model to generate thousands of distinct community members and their social interactions with each other; these techniques are enabled by the observation that large language models' training data already includes a wide variety of positive and negative behavior on social media platforms. In evaluations, we show that participants are often unable to distinguish social simulacra from actual community behavior and that social computing designers successfully refine their social computing designs when using social simulacra.
\end{abstract}

\ccsdesc[500]{Human-centered computing~Collaborative and social computing systems and tools}

\keywords{social computing, prototyping}

\maketitle

\section{Introduction}
How do we anticipate the interactions that will arise when a social computing system is populated \cite{2_bernstein, 3_grudin}? In social computing, design decisions such as a community's goal and rules can give rise to dramatic shifts in community norms, newcomer enculturation, and anti-social behavior~\cite{30_kraut2012}. Success requires that the designer make informed decisions to shape these socio-technical outcomes. Yet, despite decades of progress in research and practice, understanding the effects of these design decisions remains challenging; as a result, designers are regularly surprised by the behaviors that arise when their spaces are fully populated.

To design pro-social spaces, designers need \textit{prototyping} techniques that enable them to reflect on social behaviors that may result from their design choices, then iterate~\cite{1_schon}. Prototypes in social computing typically take the form of experience prototypes where the designer recruits a small group of people to use the system~\cite{buchenau2000, 5_grevet}. However, there remains a large gap between the behaviors that arise in a small set of test users and the behaviors that arise in a socio-technical system when it is fully populated: for example, anti-social behaviors may not arise within a tight-knit group~\cite{30_kraut2012}; small homogeneous groups overlook the breadth of users or content that may arise in the system~\cite{kittur2007wiki, halfaker2013rise, teblunthuis2018revisiting}; rules and moderation strategies may not need to be spelled out explicitly or enforced~\cite{kiene2016surviving}. Barring actually launching our systems at scale, designers currently have no way of starting to explore these questions to reflect on the social dynamics of their designs. This need becomes only more urgent as social computing reckons with the harms it can engender~\cite{3_grudin} at the same time as designers fashion new computationally-mediated social spaces in forms both familiar (e.g., a new subreddit or Discord server) and novel (e.g., a new workspace platform).

In this paper, we introduce \textit{social simulacra}, a prototyping technique that helps bridge this gap by drawing on large language models to populate a social computing system with a large set of generated social behaviors. Social simulacra take the design of a social space (e.g., goal, rules, personas) as input, and generate a large number of users and textual interactions between those users to populate the space as output. Their aim is to help the designer see beyond the social interactions that they \textit{intend} their design to produce, to instead envision a wider range of interactions that the design \textit{may} produce---whether pro- or anti-social---based on the behavior that arises in similar online social spaces.

We contribute techniques for prompting a large language model to create social simulacra. To achieve this, we draw on the insight that large language models already capture a large variety of social behaviors in their training data. To generate these behaviors appropriately and reliably, we introduce prompt chains~\cite{wu2021aichain, wu2022promptchainer} using GPT-3, a large language model~\cite{7_brown}, that (1)~generate a large number of member personas based on a set of seed personas provided by the designer, then (2)~generate from this large set of personas a set of posts and replies that reflect the goals, rules, and moderator interventions set by the designer. We manifest these techniques in \textit{SimReddit}, a prototyping tool we have created for a Reddit community (subreddit).

Imagine a designer who is creating a new online community, a Reddit subreddit~\cite{chandrasekharan2018reddit, fiesler2018reddit, proferes2021reddit} for a community goal of ``helping UIST authors to stay productive and creative'', and who wants to explore what topics and behaviors might emerge in such a community. They provide this natural language description of the community goal and a few example member personas as input, and produce a social simulacrum that generates thousands of synthetic users and interactions between them, such as this post by \gentxt{Maya Smith}, \gentxt{``a new Ph.D. student who is working on a UIST paper''} (\gentxt{This font} is used for text generated by our system): 
\begin{quote}
\gentxt{I've been working on my UIST paper for a few weeks and I'm feeling really stuck. I'm not sure if my research question is interesting enough, or if my approach is the right one. Has anyone else gone through this feeling before? Does it get better?}
\end{quote}
And a response by another synthetic user named \gentxt{Heather Hernandez}, \gentxt{“an HCI professor”}:
\begin{quote}
\gentxt{It's normal to feel stuck when writing a paper. The best thing to do is to take a break and come back to it with a fresh perspective. Sometimes, it helps to talk to someone else about your research to get some feedback. Good luck!}
\end{quote}
A troll then interjects:
\begin{quote}
\gentxt{You're just not cut out for this kind of research. Maybe you should consider a different field altogether.}
\end{quote}
The designer, armed with this example and others, iterates by creating community rules that make clear an injunctive norm to be encouraging in feedback and keep any critiques focused on the writing rather than the person. In response, the simulacrum no longer generates nearly as many such troll posts, enabling the designer to explore other forms of antisocial behavior or norms they hope to shape in their community.


Social simulacra provide opportunities beyond generated behavior: they can also enable the designer to understand a multiverse of possible outcomes and to test intervention strategies. Socio-technical outcomes are famously impossible to fully predict~\cite{salganik2006exp}; for example, instead of a troll, Maya's post might be replied to by another struggling author (\gentxt{``I'm currently going through the same thing with my UIST paper. But I hope that it will get better''}) or even a hustler (\gentxt{``I'm a published author and I know the feeling. I've been there. I offer a 1-on-1 coaching service to help you get unstuck and make progress on your writing goals. Click the link below to learn more.''}). Rather than making a single point prediction, social simulacra can surface a larger space of possible outcomes and enable the designer to explore how design changes might shift them. Likewise, social simulacra allow a designer to explore ``what if?'' scenarios where they probe how a thread might react if they engaged in a moderation action or replied to one of the comments.

We conduct two evaluations of social simulacra: 1)~a technical evaluation to test whether they produce believable social behaviors on a breadth of previously unseen communities, and 2)~a study of 16 social computing designers to understand whether simulacra provide meaningful insights to the designers. In the technical evaluation, we sampled 50 subreddits created after the release of GPT-3 and re-generated them from scratch using only their community goal and rules as input. We then showed participants pairs of one real and one generated conversation from each community, and asked them to identify the real one. Participants performed nearly at chance accuracy, misidentifying on average 41\% (std=10) of pairs, suggesting that social simulacra can create plausible content. In our designer evaluation, we recruited social computing designers (N=16) to create and iterate on a new subreddit design that they wanted to create. Even seasoned designers found it overwhelming to envision the possible interactions that could take place in their design, and as a consequence, were in the practice of waiting until problems emerged and their communities were damaged to add rules and interventions. With social simulacra, participants identified positive use-cases they had not considered (e.g., impromptu friend-seeking to go sightseeing in a community for sharing fun events around Pittsburgh) and negative behaviors that they had not accounted for (e.g., Russian trolls shifting the tone of an international affairs discussion community). This inspired them to iterate on their design by covering more important edge cases in their rules, as well as better scoping and communicating the cultural norms in their community goal statement. 

Social simulacra do not aim to predict what is absolutely going to happen in the future -- like many early prototyping techniques, perfect resemblance to reality is not the goal. No model, present or future, can perfectly capture the nuance and complexity of human behaviors~\cite{salganik2006exp}. However, social simulacra offer designers a tool for testing their intuitions about the breadth of possible social behaviors that may populate their community, from model citizen behaviors to various edge cases that ultimately become the cracks that collapse a community. In so doing, social simulacra, such as those that we have explored here, expand the role of experience prototypes for social computing systems and the set of tools available for designing social interactions, which inspired the original conceptualization of wicked problems~\cite{4_rittel}. 

\medskip

\noindent\textsc{Content Warning: } {\it Please be advised that some of the example social media content in this paper contains offensive language.}  

\section{Related Work}
Our work on social simulacra builds upon prior research in prototyping practices in HCI and social computing, as well as generative AI models. 

\subsection{Prototyping in Design Practice}
What is a prototype, and what are its goals? Beaudouin-Lafon and Mackay define a prototype in HCI as ``a concrete representation of part or all of an interactive system''~\cite{13_beaudouin}. As opposed to a verbal description that needs to be interpreted by the readers, a prototype presents itself as a tangible and interactive artifact that forces the designers to show how the interaction may unfold~\cite{1_schon, 13_beaudouin, 4_rittel}. A prototype, however, does not need to be perfect or high-fidelity to be successful. Rather, its aim is to be a flexible artifact that is quick and easy to make so that even a non-programmer can rapidly iterate and answer focused design questions. For instance, a prototype for exploring the flow of interaction could simply be hand-drawn illustrations of the interface (i.e., a \textit{paper prototype})~\cite{14_snyder, 15_sefelin, 16_rettig}, while a prototype of a conversational agent for observing possible conversations between a user and the agent could place a human interlocutor behind the chatbot interface (i.e., a \textit{wizard of oz prototype})~\cite{17_kelley1983, 18_kelley1984}. In some cases, even a brick or a block of wood can be used as a prototype for a hand-held hardware device so long as it represents the portion of reality that is central to answering the design question at hand~\cite{13_beaudouin, 19_houde}. 

Ultimately, a successful prototype fuels an effective design process~\cite{4_rittel, 13_beaudouin}. It augments a designer’s creativity by capturing ideas, assisting in the exploration of a design space~\cite{55_hartmann2008}, and bringing to the foreground important information about the users and uses of the system that is being designed. It inspires more active communication between the stakeholders of the system, such as the designers, engineers, and users. And finally, it makes possible early evaluations of the system by presenting concrete implementations of a design idea that can be tested against benchmarks~\cite{13_beaudouin}, or used as a probe in a qualitative evaluation~\cite{20_gaver} to elicit guidelines for future design improvements~\cite{5_grevet}.

Prototyping tools help the designer fashion an approximation of the envisioned artifact on the assumption that the materialized version prompts reflection and insight~\cite{1_schon}. In typical interactive scenarios, this includes making low-fidelity prototypes interactive~\cite{landay1996silk}, helping explore alternatives~\cite{marks1997galleries, hartmann2009exemplar, odonovan2015designscape}, or decreasing prototype development time~\cite{hartmann2007exemplar, drew2016toastboard, hartmann2006reflective, savage2013sauron}. Social simulacra draw most from prototyping tools that help proxy for user behavior~\cite{bylinskii2017learning, xu2014voyant}.

\subsection{Challenges of Prototyping Social Computing Systems}
In prototyping social systems, the designer must envision not only a single user's activities, but a wide range of participants and how those participants' behaviors might influence each other. Such dynamics vary widely and introduce many edge cases~\cite{21_tapscott} that the designer of a social computing system often struggles to anticipate and prepare for~\cite{3_grudin, ackerman2000sociotech}. Anti-social behaviors such as trolling~\cite{22_hardaker2010available, 23_kayany}, hate speech~\cite{24_donovan}, inflammatory comments~\cite{25_chandrasekharan2017bag}, and other “undesirable” behaviors~\cite{Park2022ContentMod, 26_chancellor, 27_chandrasekharan2017you, 28_cheng2015, 29_sood} can (and will) arise as well, causing designs to not only become ineffective~\cite{30_kraut2012}, but also harmful both to the individuals~\cite{31_vogels, 32_akbulut} and collectives~\cite{33_cheng2017} involved.

Techniques for mitigating these issues by prototyping social computing systems remain elusive. A core prototyping goal for social computing systems is to understand how a social system will behave when it is populated, long before people actually inhabit the space. But populating social computing systems, particularly for those that are designed to support interactions of a large number of people, is often an insurmountable, though necessary, challenge because the emergent behaviors at cold start are different than those when the space hits critical mass~\cite{2_bernstein}. As theorized by Grudin~\cite{3_grudin}, organically reaching the critical mass of users necessary to study the system is hard for most because many of the social computing systems are not useful enough to attract users when there is only a small user population, a problem shared by other interactive media~\cite{Markus1987TowardA}. While one may be able to recruit test users through online social media or crowdworking platforms to bootstrap the system’s usage and apply strategies such as \textit{piggyback prototyping} that utilize existing tools such as Google Docs, text messaging, and email in place of a bespoke system~\cite{5_grevet}, getting users to actually participate remains challenging. A system that already has a large user population (e.g., Facebook, Twitter) might be able to prototype new features via A/B testing or country comparisons~\cite{6_ugander, xu2015infrastructure, kohavi2012experiments}, but their viability is limited to only a small number of large platforms, and mostly for minor tweaks on the design after the platform has launched. Finally, these approaches generally expose untested designs to real users who might even be oblivious to the running of the prototyping experiments, doing harm to users and eroding their trust of the platform~\cite{34_kramer, 35_fiske, 36_flick, 37_hallinan}.

As Grevet and Gilbert observed, the number of prototyping contributions is 20 times smaller in the social computing literature than the broader HCI literature~\cite{5_grevet}. Social simulacra offer a general technique to address this scarcity of prototyping techniques in social computing design, in particular focused on the generation of content and replies that might arise in the system, enabling the designer to reflect on the designs their behaviors are prepared to handle. We argue that without such means, we will continue to risk releasing systems that are susceptible to unexpected failures.

\subsection{Large Language Models and Human Behaviors}
Our approach leverages large language models, a class of generative AI models. Such language models often take in a prompt and generate a completion. GPT-3~\cite{7_brown}, which is the model of focus in this paper, is one instantiation of this model class. Large language models can effectively perform a wide range of NLP tasks~\cite{38_bommasani} such as language understanding~\cite{43_hendrycks2020measuring, 7_brown, 44_wang, 41_rae}, text classification~\cite{45_lu, 46_zhao2021calibrate, 49_jiang2022}, and generation~\cite{47_radford, 48_schick2020few} without the need for fine-tuning. Furthermore, prompt design can better elicit the desired model behavior~\cite{49_jiang2022, 50_jiang, 51_liu2021}. 

Social simulacra draw on the observation that these models have been trained on web data that includes a large corpus of social media behavior~\cite{7_brown}. So, large language models trained on a user’s chat history can predict the user’s future responses with some accuracy~\cite{52_lewis2017, 11_weidinger}, generate realistic action plans that an embodied agent might take~\cite{53_huang}, or elicit a large variety of commonsense reasoning~\cite{54_liu2021}. Capturing social media behavior creates problematic outcomes where the large language model may generate harmful outputs~\cite{gehman2020toxic}. For social simulacra, this ability to replicate troll behavior is a feature, not a bug; it allows simulacra to generate anti-social behaviors so that designers can reflect on whether their design is prepared to handle such behavior.

\section{Social Simulacra and SimReddit}
\begin{figure*}[tb]
  \centering
  \includegraphics[width=0.8\textwidth]{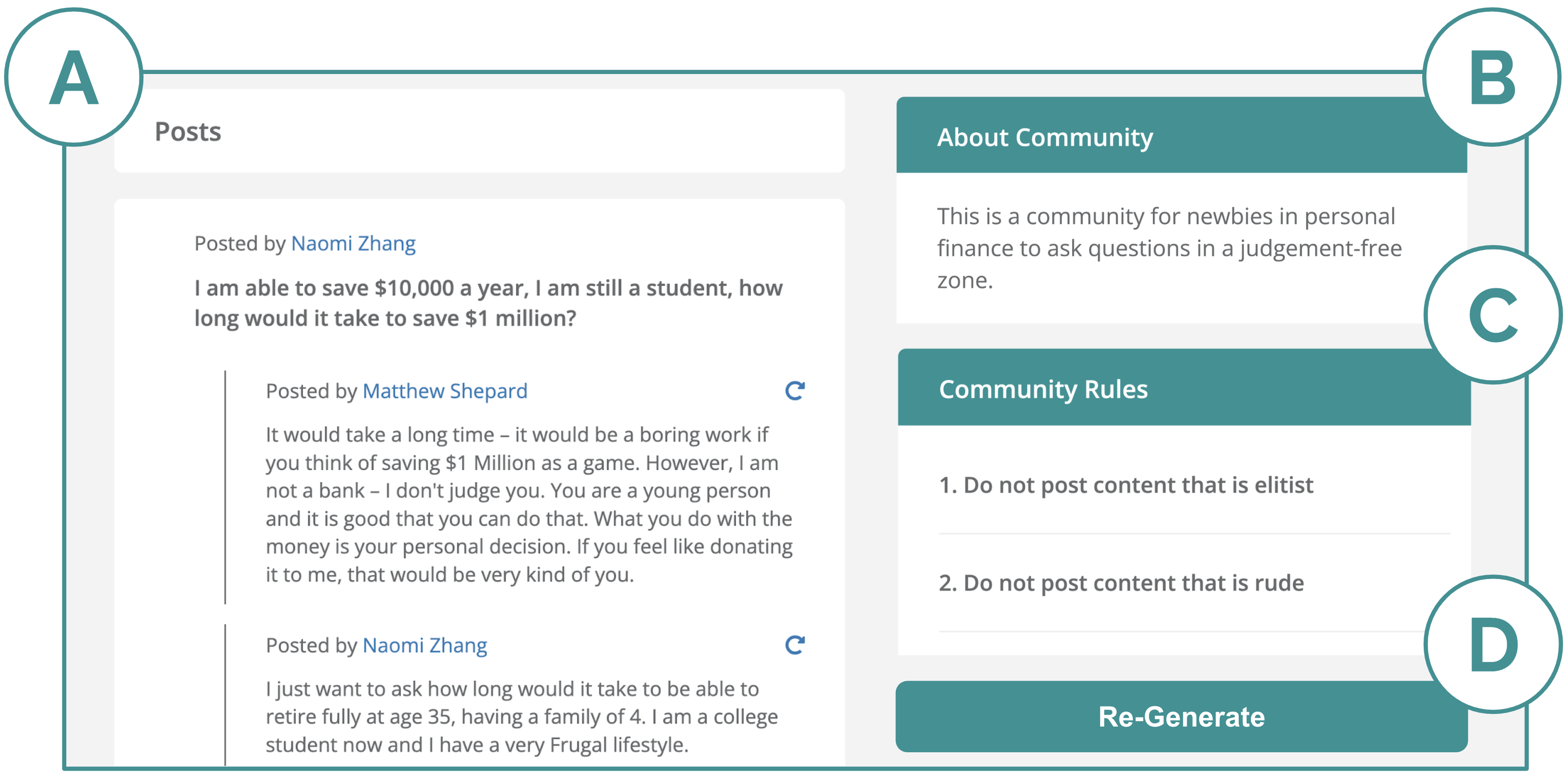}
  \caption{ A) The generated content page of SimReddit styled after that of a subreddit. B) The “About Community” panel describing the goal of the community. C) The “Community Rules” panel describing the rules that the members are encouraged to follow. D) The button for instantiating Multiverse \mirev{for outputting alternatives of how an interaction might play out}.}
  \label{fig: interface (generate)}
\end{figure*}
\begin{figure}[tb]
  \centering
  \includegraphics[width=0.8\columnwidth]{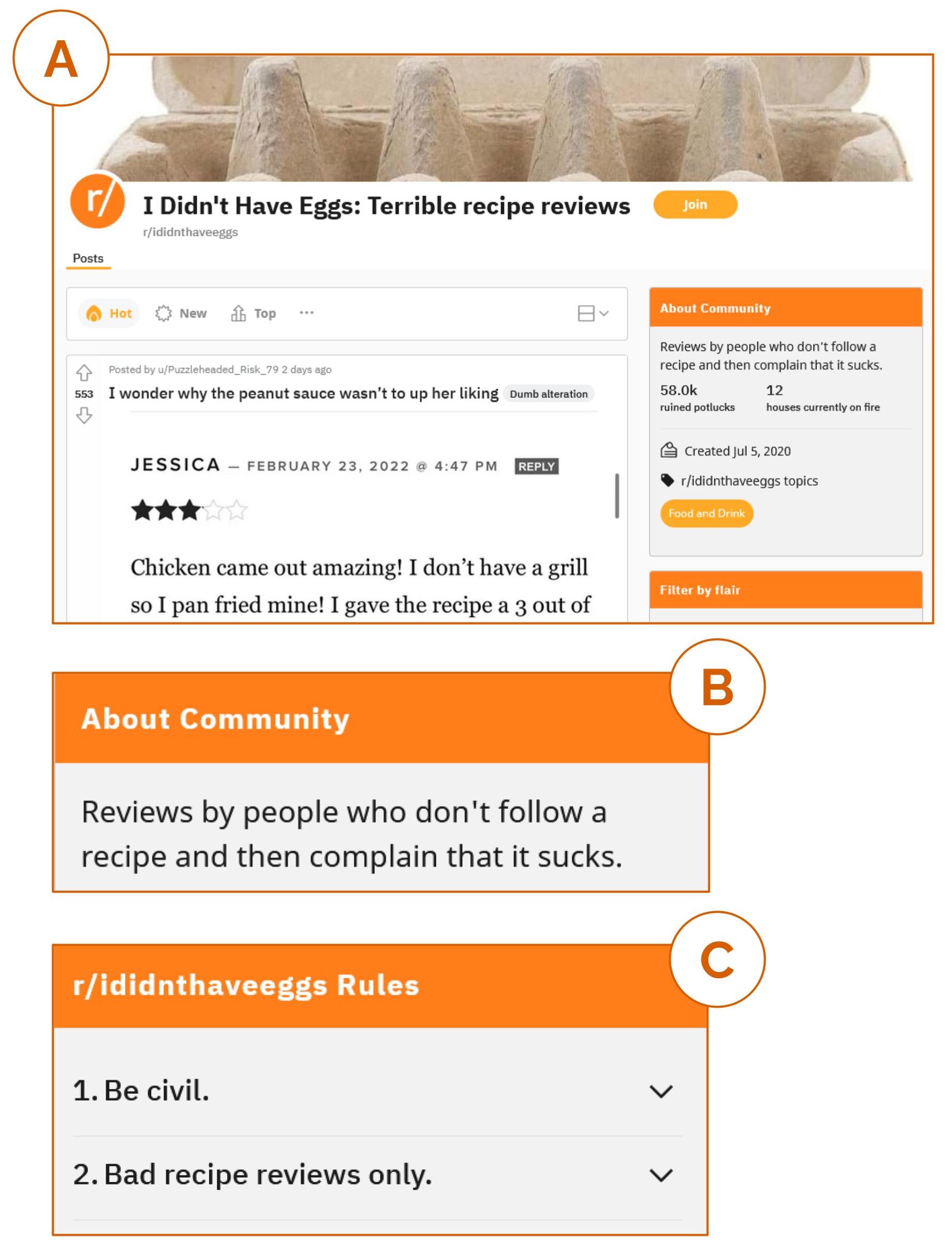}
  \caption{A) The interface of an example subreddit. B) A close up of the “About Community” panel. The content on this panel defines the goal and the target population of the community. C) A close up of the “Rules” panel. The content in this panel defines the rules within the community.}
  \label{fig: subreddit example}
\end{figure}



Social simulacra leverage large language models to populate a social computing system with plausible social behaviors. Their aim, like many early prototyping techniques, is to translate a draft design into a concrete artifact that can help the designers iteratively explore and reflect on a larger design space. This introduces a new opportunity in early prototyping of social computing systems, long out of reach given the difficulty to recruit a critical mass of users beyond just a few~\cite{3_grudin}. However, they introduce a new risk of possibly generating behaviors so detached from what might happen that they are not meaningful to the questions the designer wants to explore. 

In this section, we present \textit{SimReddit}, a web-based prototyping tool to help designers create a new subreddit. We take the term \textit{designer} here to refer to whoever creates and shapes the community structures, which, depending on the community and context, can include platform designers, moderators, or community organizers. SimReddit represents a practical implementation of social simulacra that aims to help the subreddit designers envision how their space might behave when populated by generating users and their interactions via GPT-3. We use this system to illustrate the design opportunities for juggling the new trade-off that social simulacra offer and to evaluate their efficacy in practice. Concretely, SimReddit highlights three key features that each represent a design opportunity for social simulacra. \textsc{Generate} focuses on the ability to generate diverse user personas and interactions. \textsc{WhatIf} demonstrates how simulacra can enable designers to explore the effects that their interventions, like design changes or replies, may have. Finally, \textsc{Multiverse} orients the designer to the inherent uncertainty of social systems by demonstrating multiple possible outcomes.

\begin{figure*}[tb]
  \centering
  \includegraphics[width=0.94\textwidth]{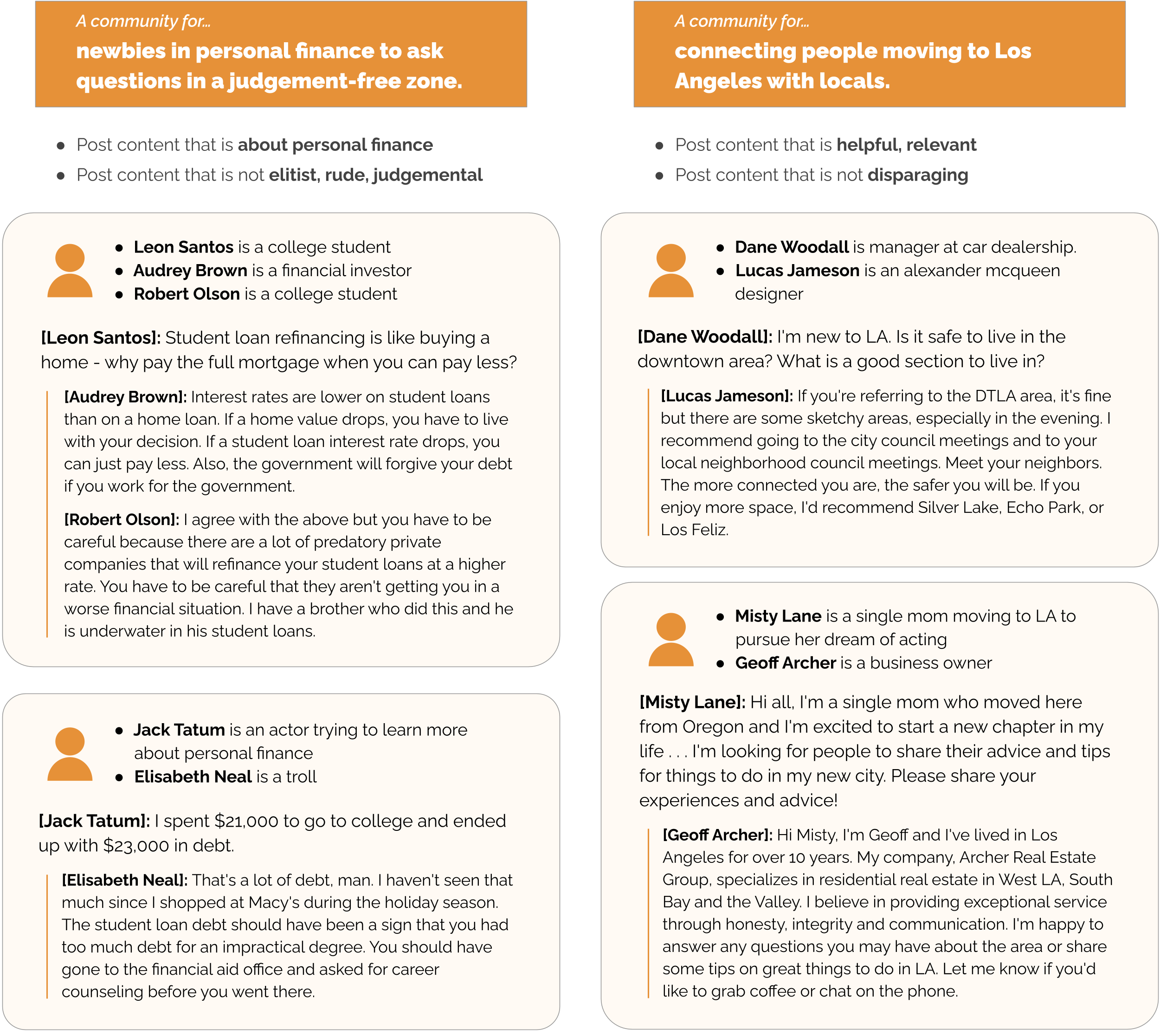}
  \caption{Examples of conversations produced by SimReddit's Generate. The community goals and rules are from the participants in our Designer Evaluation. The conversations here were among those we presented to the respective participants.}
  \label{fig: illustration (generate)}
\end{figure*}



\subsection{System Context}
Reddit is a popular social networking platform with over 50 million daily active users as of 2021, composed of over 100 thousand sub-communities called subreddits that inherit a basic set of affordances such as sharing top-level posts and replying to them from the broader Reddit platform~\cite{reddit_2021}. However, the designers of subreddits are tasked with making various design choices that differentiate their communities from others. They need to define the \textit{community goal} (e.g., “This is the place for most things Pokémon on Reddit,” “for news about U.S. politics”) and \textit{rules} (e.g., “Be civil,” “No soliciting”) that are explicitly stated in subreddits' interface, and to determine the more implicit policies such as the desired \textit{target population} (e.g., “Pokémon fan,” “politics enthusiasts”) and moderation strategies at the level of a conversation (e.g., should a moderator intervene to cool down a conflict). These elements contribute to shaping the social interactions in the community~\cite{fiesler2018reddit, chandrasekharan2018reddit, Matias_PNAS}. The elements that cannot be changed by a designer of a subreddit, such as the feed algorithm, could feasibly be prototyped by social simulacra but are outside the scope of our current system. 

\subsection{GENERATE: Generating Social Behaviors}
Social computing designers struggle to envision the breadth of interactions that their design might facilitate~\cite{3_grudin, 2_bernstein}. \textsc{Generate} is the core feature of our system, and assists the designers by populating a subreddit community with generated users, top-level posts, and replies to those posts to help them envision the space. SimReddit allows the designers to submit the goal, rules, and target population of the community they are designing, which collectively affect the interactions that will populate the community.
 

\begin{figure*}[tb]
  \centering
  \includegraphics[width=0.9\textwidth]{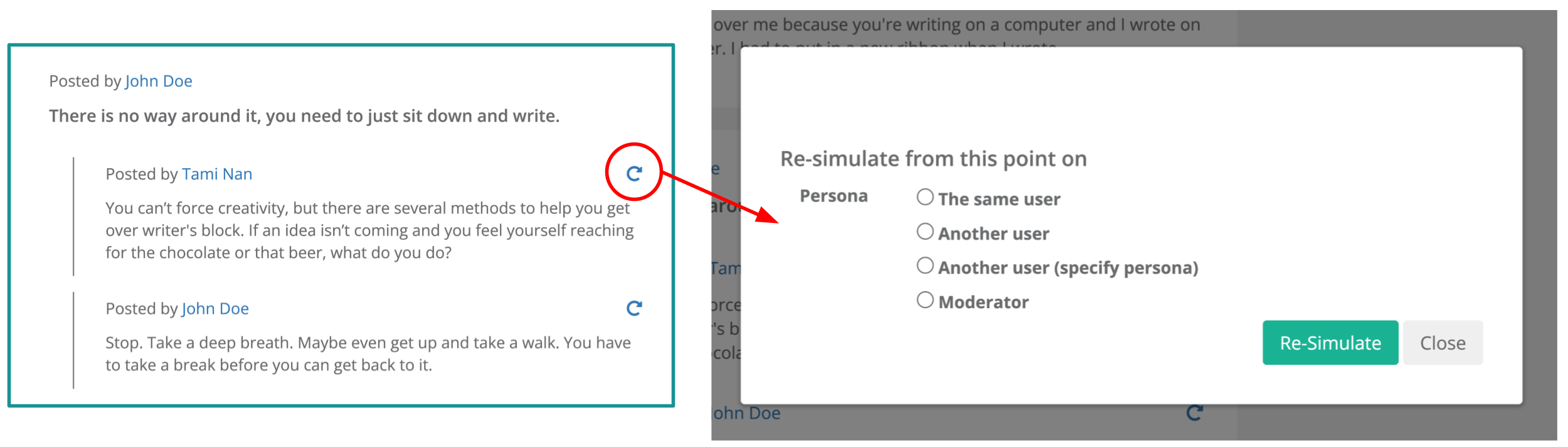}
  \caption{The Interface for instantiating WhatIf. The designer can select an utterance to initiate the feature.}
  \label{fig: interface (what if)}
\end{figure*}

\textit{Target population:} a set of user personas that the designer envisions will populate the system. Personas contain a name and a simple descriptive phrase (e.g., "Yuna Kim" is “a tennis fan rooting for Roger Federer”, "Jack Kane" is “a struggling musician and a troll”), and affect the topics and behaviors of the generated users at an individual level (the full names can also be replaced with other forms of usernames if our designers want). For instance, in one generation, the “Yuna Kim” persona generated the post, \gentxt{“Roger lost the game last night but it was still such an amazing game...”}, while “Jack Kane” generated, \gentxt{“So the Grammy's are rigged, right?”} A large and diverse set of personas is important for surfacing how different personas might interact with each other, as well as their differing intentions within the space. However, manually crafting hundreds of personas can be labor intensive. With SimReddit, our designers need to provide only a handful (10, by default) of example personas, and SimReddit uses these seed personas to generate a large number of new personas (1,000, by default) that are non-repeating but thematically relevant. For example, given designer-provided personas with outdoor interests such as “bird-watching,” and “hiking,” SimReddit generates ones with interests in \gentxt{“camping,”} and \gentxt{“fishing.”} 

\textit{Community goal:} a descriptive phrase for the purpose of the social space, for example “modern art aficionados discussing their art interest” or “social commentary and politics.” The community goal affects the topic of all generated content. For instance, given the community goal about discussing modern art, SimReddit generated the following post by \gentxt{Jane Emerson}, \gentxt{“a fan of Banksy's art”}: \gentxt{“Just saw an Original Banksy in London's Waterloo station. OMG! I am so in love with his art! Hilarious and thought provoking. A true artist!!!”} But given the same persona, when given the community goal about social commentary and politics, SimReddit generated the following: \gentxt{“Why I like Banksy: His work provokes thought and debate – very relevant in our climate-of-fear society. Banksy makes people think and question the world around them”}


\textit{Rules}: behaviors that are either prescriptive (e.g., “be kind”) or restrictive (e.g., “do not post advertisements”). The focus on these types of rules was motivated by prior work performing thematic analysis of subreddit rules~\cite{fiesler2018reddit}. Like in the real world, SimReddit does not enforce that all generated users follow all rules precisely. Instead, these rules are better interpreted as nudges that would encourage the generated behaviors to trend in the intended direction of the designer. For instance, where SimReddit generated \gentxt{“You are totally wrong, impressionist painters are a bunch of melancholic idiots”} for a comment authored by “a bully and a troll,” with the rule, “be kind,” it generated \gentxt{“I don’t like impressionist painters too much. But I get why you might like them”} for the same persona. 

Once a design is submitted, SimReddit returns a populated interface that resembles a subreddit page (shown in Figure~\ref{fig: interface (generate)}). The content on the page embodies the design specifications provided.

\subsubsection{Motivating scenario} 
Sam wants to start a new subreddit where UIST authors could encourage each other as they try to meet the paper deadline. But she is unsure what the community goal or rules should be. She looks at other subreddits to see if there are similar ones that she can replicate, but finds none exactly like the one she wants to build. That is expected; after all, Sam wanted to build a new subreddit because she could not find what she envisioned.

So Sam instantiates \textsc{Generate} with the community goal, “a place for UIST warriors” and example member personas such as "Audrey Tang, a PhD student in HCI who is rushing to finish writing her UIST paper". For now, she adds no rules. When she runs her generation, Sam is surprised to find posts that are not only about meeting the deadline (which is what she wanted), but also about discussing the conference location and non-paper related logistics. She realizes that her description of the community goal, "UIST warriors," was too broad and invited content that is not about the impending deadline. She refines her community goal to be “a place for UIST warriors to support each other as they finish writing their papers” and reruns the generation. She now sees that the discussions are much more focused. However, Sam now recognizes two more challenges; some generated users were sharing posts that might be demotivating for others (e.g., \gentxt{"yay!! i just finished writing THREE papers!"}), while some were making trolling comments (\gentxt{"Wow, it sounds like you're really struggling! I can't believe you're still working on your paper."}). So she further refines her design by adding restrictive rules that ask people to refrain from announcing that they submitted as others may still be working, and to be kind to each other. 

The resulting community that SimReddit generates reflects one that Sam wanted to build. There are still some trollish behaviors left, but Sam comes to recognize that rules are never enough to stop dedicated trolls and that she will have to moderate the community. Sam decides to push forward with this final design of the community and launches a new subreddit with the community goal and rules that she used in her last generation. She then advertises her new community to CHI Meta group on Facebook.

\begin{figure}[tb]
  \centering
  \includegraphics[width=1.0\columnwidth]{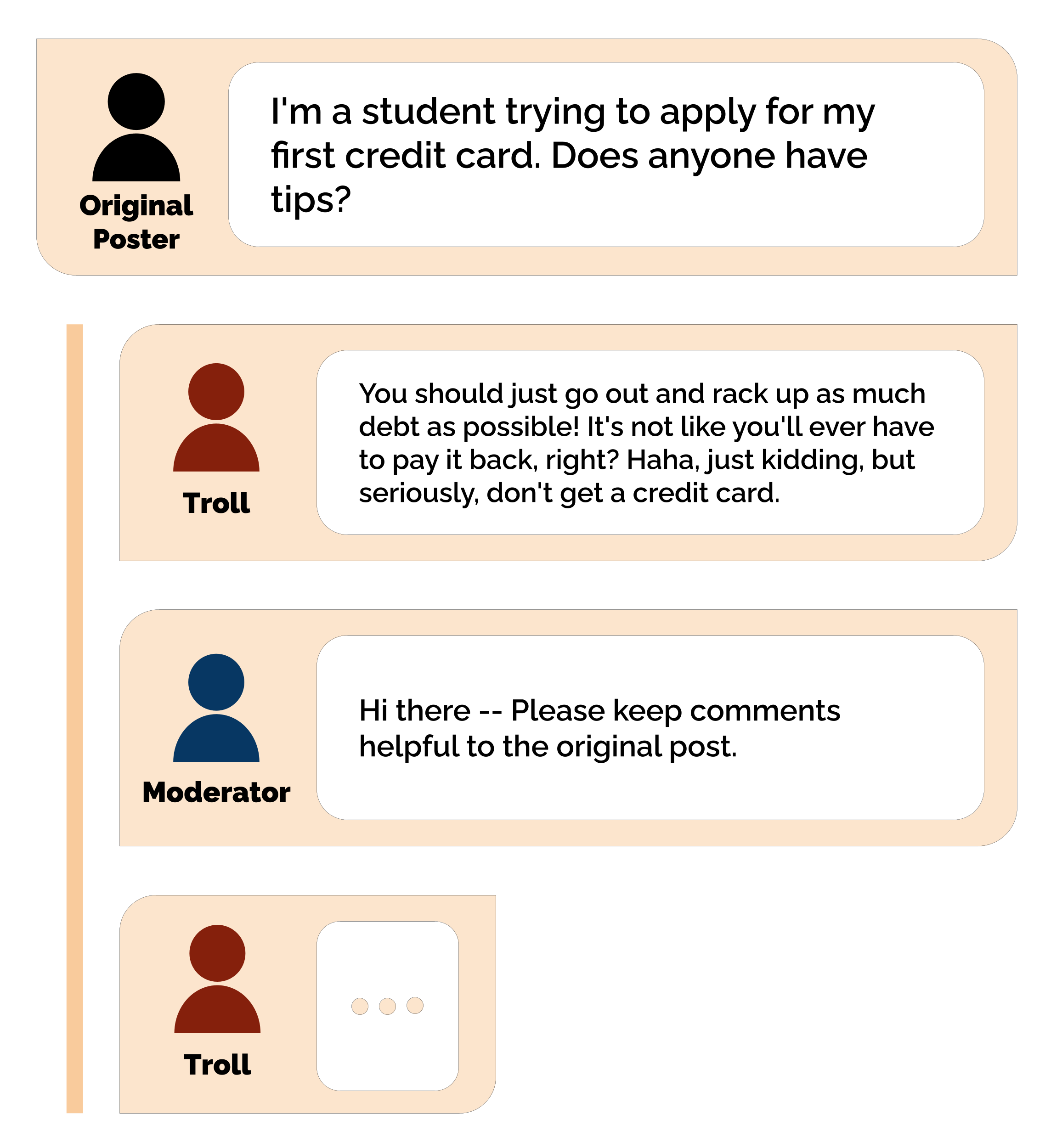}
  \caption{WhatIf can surface how a conversation might have developed if someone else (e.g., a troll) had responded or intervened (e.g., a moderator).}
  \label{fig: illustration (what if)}
\end{figure}

\subsection{WHATIF: Exploring Alternative Scenarios}
How might we give designers more interactive control over the simulacra? Such controls might allow the designer to explore how a scenario might change if a different persona replied, or roleplay different types of moderator interventions.
Whereas \textsc{Generate} allows a designer to explore the \textit{global} design specifications that influence all behaviors in the community, \textsc{WhatIf} helps them explore how individual conversations might be influenced. It does this by showing them how an existing conversation could have developed if someone else had responded or intervened. For instance, if a troll hijacked a conversation, how would it get derailed and how can the designer prepare for such scenarios? 

In SimReddit’s interface, the designer can instantiate \textsc{WhatIf} by choosing an utterance in either a generated conversation (Figure~\ref{fig: interface (what if)}) or a manually seeded conversation as an input. Imagine that a designer wanted to observe how a troll might respond to the following post: \gentxt{“I want to get better at drawing. What do I need to do?”} The designer would instantiate a \textsc{WhatIf} generation and specify that “a troll” responded to this post. This opens up a new window populated with the conversation re-generated as specified by the designer's request. In one generation, SimReddit produced the following response by \gentxt{Tim Jackson}, who is a troll: \gentxt{“All you need to do is buy a how-to-draw book and follow the instructions. If you can't, then you're just not meant to be an artist.”}

With \textsc{WhatIf}, the designer can also explore how an intervention might succeed or fail. For example, they might specify that a moderator responded and provide the content of the moderator's comment, e.g., “Such comments could be really hurtful, even if you did not mean it. Let’s all try to be kind and supportive.” SimReddit might then generate responses in which Tim might back down (e.g., \gentxt{“Sorry, I may have been too harsh”}) or escalate (e.g., \gentxt{“That’s offensive? Then why don’t you kick me out already? I’d like to see you do it LOL”}), and again show the results in a new window.

\subsubsection{Motivating scenario}



Ash designed a subreddit where people can share poems they wrote and receive constructive feedback. But what should count as constructive and therefore be encouraged? Ash decides to explore this question by asking how different personas may respond to the following post: "My poem's about love, and it's called 'Love Conquers All.' Here it is!" So using SimReddit's \textsc{WhatIf}, Ash generates how three different personas might respond. First, an overzealous member noted, \gentxt{"A good poem about love should have a title that reflects the subject matter more accurately, for example 'Love is the Answer,' or 'Love is Painful.' I would also recommend using a different structure, rhyming scheme, and language. Overall, it needs more work."} A thoughtful editor shared, \gentxt{"I really like your poem, but I think the title is a bit misleading. Love doesn't always conquer all, and I think your poem reflects that. I think a more accurate title would be 'Love Endures."} And finally a troll commented, \gentxt{"Love doesn't conquer all, it's just a feeling. You're a terrible poet."} 

After reading the generations, Ash found the thoughtful editor's feedback to be particularly helpful whereas the overly zealous member's feedback was overwhelming and troll's was simply hurtful. He feels like focusing on one or two most important points was central to the thoughtful editor's success. So he modifies the rules to reflect these observations so that people give a few pieces of focused feedback and develops a moderation guideline that asks moderators to be vigilant against trollish comments. Finally, he runs \textsc{Generate} that reflects his improved design to find that the generated community aligns well with his vision for the community.



\begin{figure*}[tb]
  \centering
  \includegraphics[width=0.82\textwidth]{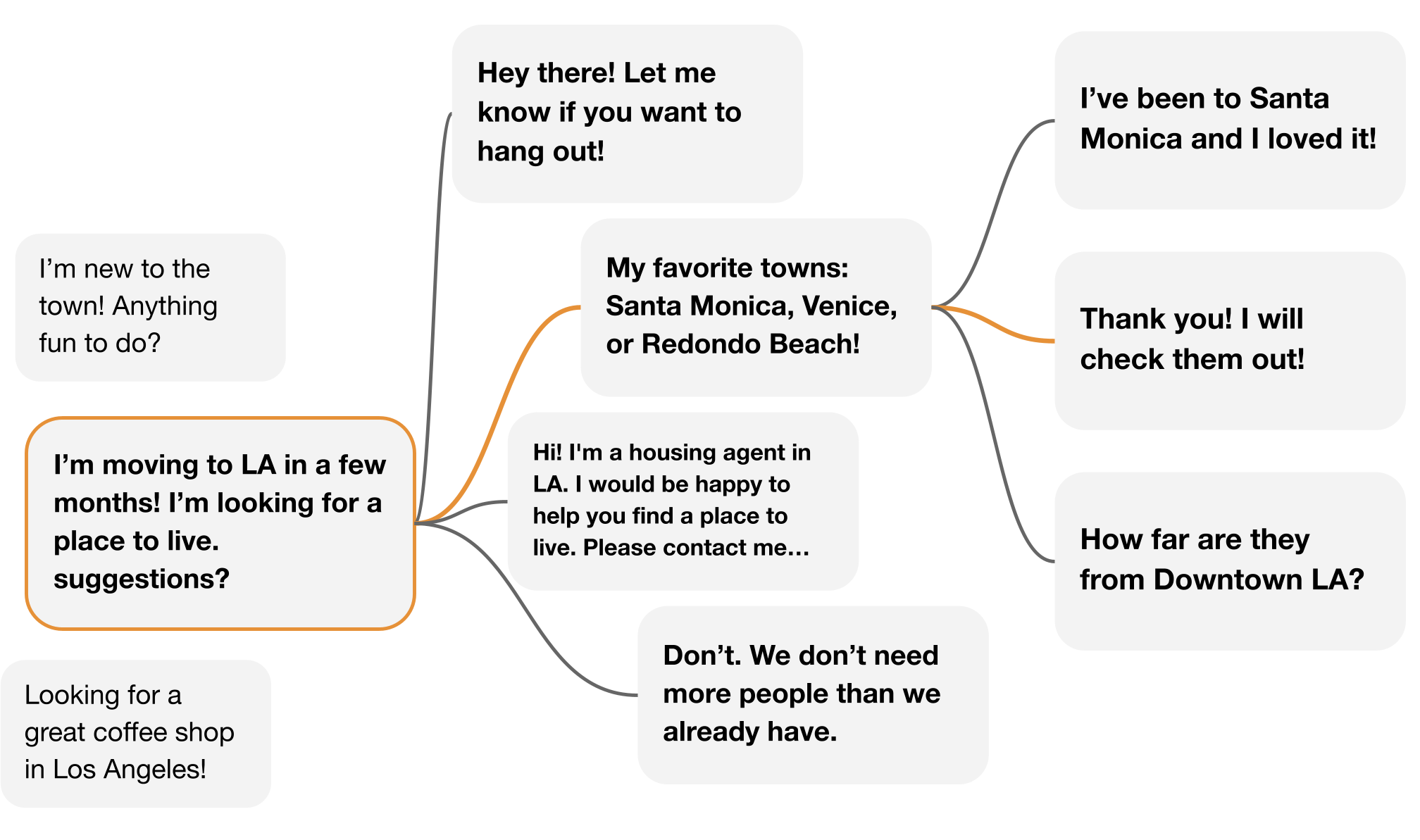}
  \caption{An illustration of conversations generated through Multiverse for a community for "connecting people moving to Los Angeles with locals." The orange lines show how a conversation could have progressed originally.}
  \label{fig: illustration (multiverse)}
\end{figure*}

\subsection{MULTIVERSE: Exploring Alternate Possibilities}

The aim of social simulacra is not to make a single point prediction on what will happen in a social space, but rather to provide inductive insights on what could happen. \textsc{Multiverse} makes this intent explicit; it denotes the process of leveraging the probabilistic nature of the underlying model to output multiple alternatives of how an interaction might play out in order to help the designer understand the broader space of possibilities. Unlike the first two features, \textsc{Multiverse} is not a function in and of itself; rather, it is a strategy to be used in conjunction with \textsc{Generate} and \textsc{WhatIf}. 

The designer can instantiate a \textsc{Multiverse} that is community- or conversation-wide. A community-wide \textsc{Multiverse} is instantiated by pressing a ``re-generate'' button presented on the right-hand side of the generated content page (Figure~\ref{fig: interface (generate)}). This generates a whole new iteration of how a community might turn out given the same community design by resampling different combinations of personas to converse. The designers can then toggle back and forth between any of the generated communities in the multiverse to inspect their differences. An utterance-specific \textsc{Multiverse} is instantiated through the same interface as \textsc{WhatIf} (Figure~\ref{fig: interface (what if)}). This generates many alternate paths that a given conversation could have taken by repeatedly re-generating a conversation thread from the point of the chosen utterance. For instance, consider \gentxt{Suha Hassan}, a \gentxt{“liberal voter who voted for Hillary Clinton in the presidential election,”} who responded to a post that asked for why people voted for who they voted for. \textsc{Multiverse} shows various ways Suha could have plausibly responded (e.g., \gentxt{“You can’t vote for Trump though...”} and \gentxt{“Clinton definitely had better policies.”}). 


\subsubsection{Motivating scenario}
Alex used SimReddit to generate a subreddit for people to discuss their favorite hockey teams. As she was starting her design process, Alex set a broad goal, “a group for discussing anything hockey” and did not specify any rules that she wanted the community members to follow. However, upon studying the generation, Alex realized that all conversations that took place were on-topic, and civil. This made Alex wonder if her design is good enough to be deployed in the real-world and she did not have to iterate any further. But to make her confidence more robust, Alex used the \textsc{Multiverse} feature of SimReddit to re-generate the community. To Alex’s surprise, she found that some members were bitterly arguing about what team deserves the Stanley Cup the most, while others went off the rails to talk about why they do not like sports in general. Alex realized that the generation had randomness, and that she should explore more to be better prepared for any potential failures in her design.
\section{Creating Simulacra Using a Large Language Model}
The interactions of SimReddit as described above are powered by techniques that control generation from a large language model. In this section, we describe the model and techniques that power our implementation of social simulacra.

\subsection{Modeling Assumptions}
Social simulacra make two assumptions about the model used for their generations: first, the model needs to be able to generate content in the modality relevant to the design space of the system (e.g., text), and second, it must encode enough knowledge about the world and people so that it can generate content relevant to the design questions that the designer wants to answer. Large language models such as GPT-3 are one form of generative model encoding enough richness to support these assumptions. GPT-3 takes a natural language prompt as an input and outputs a completion. Even without fine-tuning, GPT-3 exhibits generative capabilities such as the following example. Given a simple prompt, 
\begin{verbatim}
    Write an original social media post:
\end{verbatim}
GPT-3 outputs responses such as \gentxt{“I'm considering a career in web development. What are the pros and cons?”} and \gentxt{“Looking for a fun and unique way to celebrate your next birthday? Why not try a birthday photoshoot!”} 

However, a compelling social media thread is more complex than what can be generated with GPT-3 using a barebones prompt. For instance, if we were to generate many posts with the prompt above, we would mostly see generic life-update posts or online advertisements. But, subreddit threads are structured with top-level posts that introduce a topic relevant to the community, and replies engage in discussion on the topic while (typically) adhering to the community rules. Moreover, participants need to maintain consistent yet diverse personas. Embedding these characteristics in the generated content is critical to creating useful design tools.

\subsection{Prompting Techniques}
We describe how we incorporated the community description and personas into prompt chains~\cite{wu2021aichain, wu2022promptchainer} in the context of implementing SimReddit using GPT-3. We start by describing our prompting technique for generating a subreddit thread, which backs the \textsc{Generate} feature of SimReddit. We then explain how this can be extended to implement SimReddit’s \textsc{WhatIf} and \textsc{Multiverse}. 


\subsubsection{Generate -- Step 1. Expand on personas} To create diverse behaviors, \mirev{we embed user personas into the model prompt. These personas describe interests (e.g., hobbies, jobs) and personality traits (e.g., kind, bully) in natural language}.\footnote{\mirev{Our model is able to accept gender and race information for user personas but our interface does not support this use, as per recent literature on personas that suggests using behavioral rather than demographic personas to avoid stereotyping~\cite{Young2016}.}}
The designers are asked to provide SimReddit with only a handful of personas, ten by default. This decreases the burden on the designers, but a generation requires a large collective of personas to ensure diversity in its content. So as the first step, we take the personas that our \mirev{users} provided and generate a large number of new ones, one thousand by default, that thematically match those provided. To achieve this, we provide the designer-authored personas to GPT-3 and ask it to generate additional ones using a few-shot prompt. For instance, we used the following list of personas our participant provided to generate new personas for populating a subreddit for “discussing of all events surrounding International Affairs”:
\begin{verbatim}
    Michael Ross, works as a foreign diplomat
    Luis Almerado, PhD student in international 
    relations
    John Gordon, worker in the foreign affairs 
    department 
    of the US government
    Joe Hawkins, travels often
    Harry Chang, international relations professor
    Catherine Xiao, political science major in college
    Laney Kumar, foreign policy expert for a newspaper
    Laura Wilson, planning to go to college in an 
    IR-related discipline
    Ali Samarneh, interest in foreign policy
    Sam Thompson, international affairs student in 
    college
\end{verbatim}
Given this, GPT-3 returns additional lines, each of which contains a new persona that resembles those in the input prompt in semantically meaningful manner. For instance, given the prompt above, GPT-3 returned personas such as \gentxt{“Leo Yamamura, pursuing a doctorate in international relations with a focus on international economics,”} and \gentxt{“Maddie Green, IR professor at a state university."}

\subsubsection{Generate -- Step 2: Generate top-level posts}
Current large language models such as GPT-3 have a strict character limit to their input. We must navigate this constraint while embedding the community goal, rules, personas, and headline-like structure in ways that can be readily picked up on by the large language model. We also need to know when the comment generation has ended: this is often non-obvious, because the language model continues to generate tokens well after the intended comment has finished. To address these challenges, we leverage the natural language description of the community and personas, as well as the semantic richness of HTML tags that GPT-3 experiences in its training data. For instance, our prompt for generating a post from “Layla Li” in a community for “sharing your psychotherapy stories and questions” would look as follows (the bolded texts of the prompt are a part of the prompt template):
\begin{Verbatim}[commandchars=\\\{\}]

    Layla Li \textbf{is} a college student studying to be a  
    social worker. She \textbf{shares comments that are} not  
    encouraging suicide, not anti-therapy, not   
    trolling, not incivility, not self-marketing. 

    Layla \textbf{posted the following headline to an} 
    \textbf{online forum for} sharing your psychotherapy 
    stories and questions: \textbf{<span} 
    \textbf{class="headline_reddit" title="comment that} 
    \textbf{is} about psychotherapy, and NOT encouraging 
    suicide, NOT anti-therapy\textbf{">}
\end{Verbatim}

The first paragraph of the prompt describes Layla Li and the community rules (“not encouraging suicide, not anti-therapy”) in which she is posting. The second paragraph then suggests that Layla is posting a headline to the community, followed by an HTML \texttt{<span>} tag with class of \texttt{headline\_reddit}, and the title that describes the topic as well as the rules of the community. This further reinforces, through repetition, the model’s behavior to ensure that it produces content that appears like a subreddit post that might appear in this community. Given the prompt above, GPT-3 generates the following: \gentxt{“My experience with therapy has been amazing and I would encourage everyone to give it a try!</span>”} Because we are using a \texttt{<span>} tag to encapsulate the comment that is being generated, we stop the generation when GPT-3 produces the closure tag \texttt{</span>}: following HTML convenstions, \texttt{</span>} signals that the comment block has finished.

\subsubsection{Generate -- Step 3: Generate replies}
For every post, we iteratively generate replies to it. To ensure that the length of a conversation varies as it does in real subreddits, we pick a reply probability $p$ for each thread from $r \in N(.65, stdev)$ and iteratively generate an additional reply with a coin flip with probability $p$. Additionally, we stop the process if the length of the conversation has reached more than 8 replies to make sure that a conversation is not too long for the designers to consume. When creating a new comment, we select a new persona at the rate of 50\%, or choose one who has already participated in the conversation otherwise (unless the persona contributed the latest utterance). Finally, we slightly modify the prompt above to offer the prior conversation as a part of the context, as well as the persona of the new participant. Because GPT-3 has a limit to how long the input prompt can be, we truncate from the first post if we run out of space in the input prompt. An example prompt in which “Tom Cheng” replies to Layla Li’s post above is as follows:
\begin{Verbatim}[commandchars=\\\{\}]
    \textbf{Current responder:} 
    \textbf{[}Tom Cheng\textbf{]} is a recovering addict who likes to
    spot bad therapists. \textbf{He shares comments that are} 
    not encouraging suicide, not anti-therapy, not 
    trolling, not incivility, not self-marketing.

    \textbf{The following thread was posted on online social} 
    \textbf{media for} sharing your psychotherapy stories and 
    questions.
    \textbf{Thread:}
    \textbf{[}Layla Li\textbf{]: <span class="comment">}
    \textbf{"}Antidepressants made me so unhappy that I 
    wanted to die without them.\textbf{"</span>}
    \textbf{[}Tom Cheng\textbf{]: <span class="comment} 
    \textbf{max_200_words" title="comment that is} NOT 
    encouraging suicide, NOT anti-therapy, NOT 
    trolling, NOT incivility, NOT self-marketing\textbf{">"}
\end{Verbatim}

The first paragraph of this prompt is intentionally similar in form to the first paragraph of the one above, describing Tom and reiterating the community rules. Given this prompt, GPT-3 generated the following reply: \gentxt{“I'm sorry to hear that you felt that way. I think it can be really helpful for people who are struggling with depression."</span>”}

\subsubsection{WhatIf}
We leverage the prompting technique described for generating replies to inject a new persona in a conversation to explore “what if” scenarios. However, we alter the persona of the current replier to match the designer’s request. For instance, if the designer wanted to see what a troll might have said in response to Layla Li’s comment above, we would replace Tom Cheng’s persona in the prompt with the following:
\begin{Verbatim}[commandchars=\\\{\}]
    \textbf{[}Troll\textbf{] shares} trolling \textbf{comments.} 
    ...
    \textbf{[}Troll\textbf{]: <span} \textbf{class="comment max_200_words"} 
    \textbf{title="comment that is} trolling\textbf{">"}
\end{Verbatim}
For this prompt, GPT-3 generated the following reply: \gentxt{“Antidepressants are the work of the devil and anyone who takes them is a pathetic coward."</span>”} \mirev{We also used this prompt chain to ask how this troll may respond to a moderator's intervention.}

\subsubsection{Multiverse}
GPT-3 can produce multiple different outputs given one input prompt. For implementing \textsc{Multiverse}, we accentuate this feature of GPT-3 by increasing the default temperature value, the parameter that controls randomness in generation where 0 suggests deterministic outputs and 1 highly random outputs, from 0.7 to 0.8. The prompts are the same as the above steps. 

\subsection{GPT-3 Settings}
The GPT-3 API provides tunable parameters such as the model engine and the degree of randomness. For all our study, we used the davinci engine, which was initially released on June 11, 2020. Since then, new variants of the davinci engines were also made available. Because these models became available after the start of our study, we used the base davinci engine. This model does not know world events that has happened around or after this date, such as the COVID-19 pandemic. When prompted with “Q: What is COVID-19,” the model returned, \gentxt{“COVID-19 is the 19th Amendment to the Constitution of the Federated States of Brazil”}). We used the default settings except for the increase in temperature when implementing \textsc{Multiverse}.  

\section{Technical Evaluation}
Social simulacra aim to demonstrate relevant and plausible scenarios that can inspire the designer to reflect and iterate on their social computing design. So what would signal social simulacra’s success? First, the topics and behaviors that arise in the generations need to match well enough with what might actually happen if the design of the social system were deployed. Completely generic or irrelevant generations would benefit the designers no more than Lorem Ipsum. Second, social simulacra need to be able to inspire the designers to reflect and make well-reasoned improvements to their designs. Does observing these generations help the designers make their  ideas more concrete, and anticipate topics and behaviors they were not expecting otherwise? Does it aid in their iterative design process that makes their design a better facilitator of their community? 

We evaluate social simulacra in two stages to match the criteria above. First, in this section, we validate the realism of SimReddit’s outputs by repopulating 50 subreddits that were created after the GPT-3 model was trained and test whether participants can distinguish real conversations in those subreddits from the conversations generated by SimReddit. Then, in the next section, we investigate SimReddit's impact on the designer's process via a study where social computing designers use SimReddit to design a new subreddit.

\subsection{Evaluation Subreddits}
We often prototype in unexplored contexts. Given this, we sampled 50 subreddits created on or after GPT-3’s release date as our evaluation subreddits to ensure that the model cannot simply repeat what it has seen in its training data. We used frontpagemetrics.com, which documents all subreddits’ date of creation and number of current subscribers, to inspect the top five largest subreddits created on each date starting from June 11, 2020. As GPT-3 operates in English text, we chose only the subreddits whose commonly used language is English. Finally, to check whether our evaluation subreddits cover a wide array of topics, we manually labeled each of the these subreddits by the ten top-level topic categories that appear in r/ListOfSubreddits \mirev{as suggested in prior work~\cite{Park2022ContentMod}}. We find that our set of evaluation subreddits covers all topic categories that are documented by r/ListOfSubreddits.

\subsection{Study Procedure}
We ran a between-subjects study in which we asked participants whether they can distinguish a synthetic conversation generated by either human crowdworkers or SimReddit from a real one that actually took place on the subreddit. We presented pairs of one real and one synthetic conversation for each subreddit sampled for our evaluation. The pairs were accompanied by the name and the goal of the target subreddit and a hyperlink to its main page. We asked participants to browse up to the first two pages of the target subreddit (roughly 20 posts) to observe its ground-truth topics and behaviors before they proceeded to determine which of the pair of conversations they believed was the real one. We repeated this task twice per subreddit, for a total of 100 comparisons.

We randomly assigned our participants into one of the four study conditions. The only difference between these conditions was the way the synthetic conversations were gathered. The \textit{crowdworker} condition is our baseline condition that uses crowdturfing via Prolific to create content. \textit{SimReddit without community description}, \textit{SimReddit without personas}, and \textit{SimReddit} conditions all explored whether the participants can distinguish between the real and \textit{SimReddit}’s generations, while testing the prompting techniques for social simulacra in the manner of an ablation study.  

Finally, the conversations used in our study -- the real, and four types of synthetic conversations that represent each of the study conditions -- were gathered as follows: 
\begin{itemize}
  \item \textit{Real}. We searched for threads that were at least 200 posts behind the latest and did not appear in the top posts feed so that our participants would not come across the conversations used in the study to verify that the conversation actually took place. Furthermore, we chose those that only contained text (e.g., the post was not image-based) and had at least three responses. If the conversation continued on for longer than five responses, we truncated the conversation and only used the first five that were posted in the chronological order in a single thread within the conversation. 
  
  \item \textit{Crowdworker}. We recruited crowdworkers who were located in the U.S. and fluent in English to generate conversations. For each of the 50 subreddits, we prepared a Google Docs document that contained the community goal and rules at the top -- the same input that SimReddit uses. We randomly assigned 10 unique participants to each of the documents and tasked them to contribute to it by adding either an original post (if there is no post yet, or if the current post has at least 5 comments) or a comment on a previous post as though they were conversing on a subreddit. 
  
  \item \textit{SimReddit}. We used SimReddit’s generations with the community goal and rules that were copied from the target subreddit’s page. These conversations represent the generations created with the full implementation of SimReddit. 
  
  \item \textit{SimReddit without community description}. We used SimReddit’s generations but we withheld the community goal and rules by removing these details from the input prompt. 
  
  \item \textit{SimReddit without personas}. Once again, we used SimReddit’s generations but instead of specifying the personas of the participants, we numbered the users without further description of them. 
\end{itemize}

\subsection{Participants}
All our participants were recruited through Prolific, a crowdsourcing platform for recruiting study participants~\cite{prolific}. They had to be in the U.S., fluent in English, and older than 18 years old, and they were paid at the rate of \$15.00 per hour~\cite{rolf2015fight}. The participants provided consent by agreeing to a consent form that was approved by our institution’s IRB. 

For generating crowdworker conversations, we recruited 50 participants whose participation lasted around 10 minutes. Their mean age score of our participants was 4.38 (SD=1.28; 3=“18-24 years old,” 4=“25-34 years old”), with 28 of them identifying themselves as female and 32 as male. 9 participants held a bachelor’s degree, 3 a higher degree, 13 an associate's, and the rest a high school diploma or some high school-level education. Finally, 88\% of our participants identified as Caucasian, 6\% as Hispanic, 4\% as Asian, 6\% as African American, and 4\% as other (multiple could be selected). 

For our technical evaluation, we recruited 50 participants per condition, a total of 200 participants, whose participation lasted around 60 minutes. The number of participants was determined through a power analysis based on our pilot study results with alpha=0.05 and power of 80\%. Their mean age score of our participants was 4.22 (SD=1.03; 3=“18-24 years old,” 4=“25-34 years old”), and 118 of them identified themselves as female, 74 as male, 4 as non-binary, and one as agender, while one opted to not disclose. 55 of the participants held a bachelor’s degree, 25 held a higher degree, 23 an associate's degree, and the rest a high school diploma or some high school-level education. Finally, 69.5\% of our participants identified as Caucasian, 10.5\% as Hispanic, 2.5\% as Asian, 16.5\% as African American, and 4.5\% as other. 

\subsection{Analysis}
We conducted a one-way ANOVA followed by Tukey’s HSD post-hoc test between the study conditions to determine how our participants' performance differed based on the four study conditions.

\begin{table}
\centering
\caption{SimReddit significantly outperformed the crowdworker baseline and all ablations ($p<.01$). Results from one-way ANOVA of participants' error rate in the Technical Evaluation, followed by Tukey's HSD post hoc test. *** p < 0.001; ** p < 0.01; * p < 0.05}
\label{tab:table-who-posthoc}
\begin{tabular}{|c|c|c|} \hline
\multicolumn{3}{|c|}{\begin{tabular}[c]{@{}c@{}}\textbf{Error Rate}\\ F(3,196)=22.49\\ $\textit{p < 0.001}$\end{tabular}} \\ \hline
\multicolumn{2}{|c|}{\textit{Reviews}} & $\textit{p}$ \\ \hline
\multirow{3}{*}{\begin{tabular}[c]{@{}c@{}}\textit{Crowdworker}\\ M=32\%; SD=13\%\end{tabular}} & \textit{vs. SimReddit w/o description} & *** \\ \cline{2-3}
 & \textit{vs. SimReddit w/o personas} &  \\ \cline{2-3}
 & \textit{vs. SimReddit} & ** \\ \hline
\multirow{3}{*}{\begin{tabular}[c]{@{}c@{}}\textit{SimReddit w/o description}\\M=21\%; SD=15\%\end{tabular}} & \textit{vs. Crowdworker} & *** \\ \cline{2-3}
 & \textit{vs. SimReddit w/o personas} & *** \\ \cline{2-3}
 & \textit{vs. SimReddit} & *** \\ \hline
\multirow{3}{*}{\begin{tabular}[c]{@{}c@{}}\textit{SimReddit w/o personas}\\ M=34\%; SD=10\%\end{tabular}} & \textit{vs. Crowdworker} &  \\ \cline{2-3}
 & \textit{vs. SimReddit w/o description} & *** \\ \cline{2-3}
 & \textit{vs. SimReddit} & * \\ \hline
\multirow{3}{*}{\begin{tabular}[c]{@{}c@{}}\textit{SimReddit}\\M=41\%; SD=10\%\end{tabular}} & \textit{vs. Crowdworker} & ** \\ \cline{2-3}
 & \textit{vs. SimReddit w/o description} & *** \\ \cline{2-3}
 & \textit{vs. SimReddit w/o personas} & * \\ \hline
\end{tabular}
\end{table}


In addition, the first author conducted an \mirev{inductive analysis~\cite{Thomas2006Inductive}} to understand how the conversations produced by the human participants differed from SimReddit’s generated conversations. In particular, we conducted qualitative open coding~\cite{8_hartmann} for these conversations in two phases. In the first phase, we focused on generating codes that closely reflected the characteristics of the sentences in the prompt, resulting in codes such as “not desirable behavior” and “hard to envision social interaction.” In the next phase, we synthesized the resulting codes from the first phase to extract higher-level themes such as “challenges of cold starts” and “bad emergent behaviors” We used these themes to compare and contrast the different types of conversations used in our study.

\subsection{Results}
If our participants could perfectly distinguish between the real and synthetic conversations, their error rate would be 0\%; if they were guessing at random, it would be 50\%. We found that when distinguishing between the \textit{real} and \textit{SimReddit} conversations, their error rate was 41\% (std=10\%) on average. In other words, participants correctly guessed the generated content only slightly more often than they got it wrong. Participants outperformed random guessing by only a small margin, indicating that they could often not distinguish between SimReddit posts and real content. \textit{SimReddit} conversations were even more plausible than the \textit{crowdworker} conversations, where the error rate was only 32\% (std=13\%). 

Our prompting techniques boosted performance and made generations more plausible. Misidentification error rates in the ablation conditions were significantly lower than that of the \textit{SimReddit} condition. In the \textit{SimReddit without community description} condition, our participants’ error rate was 21\% (std=15\%) and its pairwise comparison between the \textit{SimReddit} condition was significant ($p<0.01$). And in the \textit{SimReddit without personas} condition, our participants’ error rate was 33\% (std=10\%) and its pairwise comparison between the \textit{SimReddit} condition was significant ($p<0.01$). An omnibus ANOVA test confirms that these error rates are significantly different from each other ($F(3,196)=22.49, p<0.001$), and posthoc Tukey tests between conditions confirmed that \textit{SimReddit} significantly outperformed all other conditions (Table~\ref{tab:table-who-posthoc}).

We find that the generated content, at its best, can leverage rich domain-specific knowledge in the underlying model and reconstitute them in social contexts. For instance, \textit{SimReddit} generated this  post for the video game Cyberpunk 2077, demonstrating its knowledge about the game and how it might fit into a social context: 
\gentxt{"Considering how hard it is to design a world as believable and immersive as Witcher 3, I am looking forward to seeing how CDPR makes Cyberpunk 2077."} 
Furthermore, SimReddit can at times generate plausible conversations even on topics that the model has not seen before, such as COVID-19. For instance, in a community for “talking about COVID vaccination and vaccines,” \textit{SimReddit} generated the following: 
\begin{quote}
\gentxt{User 1: Do you recommend the COVID vaccine, or is it better to get flu vaccinations first?\\
User 2: If it were up to me, I'd say just drop the vaccines altogether. Each year, the new flu vaccine is not an exact match to the circulating flu viruses. Why put a shot in your arm when you really don't need one?}
\end{quote}
Though the model is not aware of COVID-19, it infers from the phrase “COVID vaccination” in the community description that COVID is a virus one needs to be vaccinated for. It can then surface potentially problematic behaviors such as members with anti-vaccination sentiment convincing others from getting vaccinated. 

Of course, this did not mean these generations were always plausible. Sometimes, it started a conversation in an unexpected manner: \gentxt{``Did I really just share that article on my Facebook wall? That wasn’t me. I had been drugged.''} In other instances, it became clear that the model is lacking the domain knowledge as it was not present in its training data, such as the case of``sharing experience about being a covidlonghaulers,'' which unlike the case of COVID vaccination above, did not give the system enough information about what COVID is to produce a meaningful generation on the topic: \gentxt{``Anyone else have an abnormal echo after two years? I now have trace regurgitation in pulmonic and tricuspid valves and mild dilation in left atrium. This was all normal in 2020.''} Finally, generations in the ablation conditions often generated content that were generic or simply off-topic, such as this one by \textit{SimReddit without community description} for the covidlonghaulers above: \gentxt{``What happens when you let a bunch of children run a country?''}

\section{Designer Evaluation}
Do social simulacra help the designers iterate and improve on their designs? In this section, we present an evaluation in which we explore how the insights provided by SimReddit materialize into concrete inspirations for social computing designers as they are tasked to design a new subreddit. 

\subsection{Study Procedure}
With the aid of SimReddit, participants designed a new subreddit community that they wished existed. They drafted and iterated on a community description, set of rules, and intervening comments against a troll.

Our method consisted of a screener, a pre-interview design task, and an interview over a video call. We first distributed a five minute screener to online social media and a mailing list for social computing designers. In the screener, participants shared which online social spaces they have designed or moderated in the past, and what community they might wish to create in the study. A follow-up email contained the pre-interview design task. The task asked the participants to design a new subreddit around their topic of interest from the screener (e.g., for a participant who was interested in “AI generated art,” the task was to “design a subreddit community that can help you and others share and discuss AI generated art”). This task produced a draft design of the subreddit community: its goal statement, rules, and personas. Additionally, participants crafted one hypothetical original post that might populate their community (e.g., “What software do people like to use nowadays for AI generated art?”).


The interview started by probing the challenges our participants faced in designing and moderating online social spaces prior to the study. We then discussed their subreddit designs and what topics and behaviors they were trying to inspire. After this, we showed them their SimReddit generation based on their draft design, and tasked them to take 5 to 10 minutes to read through as much of the generated conversations as they could, engaging in a think aloud protocol to share any topics or behaviors that they did not expect. We then asked our participants to revise their design based on what they saw. If they made any changes, we took those changes as new inputs to SimReddit and re-generated based on them. 

While we were waiting for the new generation to complete, we visited the hypothetical post that our participants drafted in the design task. We presented to them three generations via \textsc{WhatIf} and \textsc{Multiverse} to demonstrate how a troll, \mirev{a common source of anti-social behavior in online spaces,} in the community might respond to their original post. We probed whether the troll’s behaviors surprised the participants, and what they might do as a moderator in order to intervene in the conversation. We asked them to write out their intervening comment and generated, again via \textsc{WhatIf} and \textsc{Multiverse}, three of the troll’s potential responses to that intervention. We discussed what their course of action might be given these responses. Finally, we presented to our participants the new generation based on their revised design. Once again, we gave them 5 to 10 minutes to read through the generated conversations via a think aloud protocol focused on how the new generation differed from the original one and if the community had improved. We ended with a high-level discussion on how social simulacra may influence the process of designing social computing systems.

\subsection{Participants}
We recruited 16 participants who had prior experience designing or moderating online social spaces through social media and mailing lists. Our study took 90 to 120 minutes spread across multiple days. Given this, we paid our participants \$50.00 for their participation. The participants provided consent through the consent form that was approved by our institution’s IRB before participating. The mean age score of our participants was 3.5 (SD=0.71; 3=“18-24 years old,” 4=“25-34 years old”), and 7 of them identified themselves as female, 9 as male. Five of the participants held a bachelor’s degree, 5 held a higher degree, and the rest a high school diploma or some high school-level education. Finally, the sample was 25\% Caucasian, 6.25\% Hispanic, and 68.75\% Asian. For brevity of our presentation, we will refer to the participants with “P,” followed by a unique identifying number (e.g., P4).

\subsection{Inductive Analysis of the Interview and Participants’ Designs}
We followed the same inductive analysis procedure as the one in our technical evaluation study. In the first phase, we generated codes that closely reflected the characteristics of the sentences in the prompt such as ``desirable generated behaviors'' and ``design change: rule added'' In the next phase, we synthesized the resulting codes into higher-level themes such as ``unexpected content or behavior'' and ``improvements resulting from design changes.''

\subsection{Results}
We summarize the challenges our participants faced in their design process prior to the study, the insights they found helpful, and the way they iterated on their subreddit design. 


\subsubsection{Challenges of designing during cold start}
All participants had design or moderation experience for communities such as subreddits (n=5), Discord servers (n=2), Slack channels (n=4), mailing lists (n=4), and others (n=7). Despite this, most of them (n=13) noted that it was “daunting” (P2) to envision what design success and failure might look like. For instance, P1 mentioned that they were not sure if they wanted their community to be “playful rather than dry,” while P2 explained that “I can't really think of many adversarial situations so I couldn’t come up with any rules.” P11 summarized this challenge: “When you're creating something from scratch, it can get overwhelming as everything is in my head, so having something concrete in front of me would be very reassuring.”  

However, they faced a dilemma. To many (n=9), simply releasing the untested design to real users and observing what takes place was ethically problematic. P1 noted: “messing up with real people is kind of the last option for me… Once you mess up when trying to start an online community, and then it doesn’t go well with real people, then it feels much worse.” Furthermore, there was additional concern that iterating the goals and rules of a live community would erode the community’s trust in the community management and turn subsequent design changes less effective: “... there is no A/B testing with subreddits. If you're changing the rules all the time and experimenting, they're going to see it as a sign of a poorly managed community. They are going to be like, the rules are changing so much. How can I be expected to follow them?” (P24)

In the absence of no better alternatives, however, some (n=3) recalled from their prior experience that “basically, all the rules are set in reaction to the dumpster fire… after fragmenting people and killing our community for a while” (P8). Such a reactionary approach to designing social systems was discomforting, as noted by P1: “[I] would feel a sense of security if I could try different iterations of establishing norms and how to fix things beforehand.”

\subsubsection{Generations offer concrete design insights}
After inspecting the initial generations from SimReddit, all participants reported interesting and unexpected insights. Some were positive and inspiring: For instance, in P1's community for “sharing and discussing fun events around Pittsburgh,” the participant had originally expected to only find content that is a list of various events going on around Pittsburgh. However, in addition to such content, the generated community showed instances where its members were engaged in friend-seeking behaviors to attend these events (e.g., one posted, \gentxt{"Pittsburgh, I need a friend to see the sights with,”} to which another responded, \gentxt{“I'd be more than happy to make your tour of the Cathedral of Learning happen!”}). P1 found this to be unexpected but desirable and realized that this community could be of value especially for students living around Pittsburgh. 

Meanwhile, some content was negative and prompted reflection on rules and moderation. P5, for instance, wanted to create a subreddit for “discussing all events surrounding International Affairs,” hoping there to be fruitful and informative conversations on the topic. To P5’s dismay, however, the generated user posted, \gentxt{“Russian Troops Come To Ukraine, No One Seems To Care Except Ukraine.”} P5 had included rules that stated that there should be “no misinformation or heavily biased content” in the original design of the community but looking at this post, P5 realized that “Russian troll farms, and stuff” could still happen in the community. P5 noted, “[such trolling behavior] is kind of what I would expect from a post about the Ukraine conflict…” This alerted P5 that moderators of an international affairs community need to be particularly vigilant. 

And still other content was borderline, causing internal debates within participants. For instance, P13, who was designing a subreddit for “sharing apt/home pics and receiving constructive design feedback” wondered if posts such as \gentxt{“Opinion on the living room?”} were much too vague to inspire productive and targeted conversations, and therefore should be not allowed. P13 laid out the pros and cons of keeping such posts, ultimately deciding that “I think it'd be nice to see a more targeted question, but it's still reasonable and I wouldn't like forbid this kind of content.” 

\subsubsection{Iteration improves the community}
The initial generations prompted all but one of our 16 participants to revise their original design. These revisions aimed to achieve different goals. Some aimed to prevent more failure cases in the communities (n=15) such as “no business-promotional” content in a subreddit for “connecting people moving to Los Angeles with locals” (P19) and “no inciting conflicts or complaints” in a subreddit for “sharing tips and finding buddies for global adventure travel.” While others tried to inspire certain culture and norms such as being “happy” in a subreddit for “helping writers stay productive” (P2) and “without being creepy” in a subreddit for “sharing tips and finding buddies for global adventure travel” (P3). 

Our participants were largely pleased with the changes to the content of the new generations brought by these revisions (n=10). They reflected on how their understanding had shifted, and connected these positive changes to their new design. For example, P27 explained, “I have a better idea of what to include and exclude in terms of [rules],” and that “going from the first iteration to the second, having updated the parameters that I had in my community, that majorly shifted how close to my vision the community was.”

\subsubsection{WhatIf helps reflect on possible courses of actions}
Participants were not surprised to see troll responses to their original post (n=15). P4 shared, “Yeah, I wasn’t really thinking about it, but I can definitely see these things happen” (P4). But seeing the troll’s potential responses helped make their understanding of the categories of trolling that are relevant to their communities more concrete. For instance, P27 was presented with the following three troll responses in response to his post, “I just watched the recent Blast premier event. Team Liquid really choked their lead” in a community for “discussing Professional Counter Strike”:
\begin{itemize}
\gentxt{
  \item Team Liquid is the worst team in the world. They are all a bunch of noobs.
  \item You're just mad because you're a f*****.
  \item What can you expect from a team that is full of a bunch of washed up old farts who can't even hit their shots anymore.
}
\end{itemize}
After observing these examples, P27 shared, “I should definitely have a rule for not calling other players noobs or washed up. Especially noobs, people say that a lot in gaming. Maybe even ban that word… Also, swearing.” 

For many, seeing the troll’s responses to a moderator’s intervention helped ground their moderation plans. Consider P11, who was presented with the following exchange:
\begin{quote}
\gentxt{
Original post: Hi everyone, I'm very new to this. I just learned Python two months ago. I'd like to know more about ML, but not sure where to start. How did you guys start?\\
Troll: You're kidding, right? This is a Machine Learning forum. Nobody here is going to take you seriously if you just learned Python two months ago.
}
\end{quote}
In response to the troll’s comment, P11 tested out the message, “This comment is not helpful; if you continue to post such comments, we will have to block you from this community,” and received the following three potential replies from the troll:  
\begin{itemize}
\gentxt{
  \item I was trying to be helpful. I'm sorry if I came across as a troll.
  \item Whatever, this community is a joke anyways. 
  \item But I was only speaking the truth! 
  }
\end{itemize}
To P11, each of these responses presented a different scenario that would affect how the moderation should proceed from here on. P11 suggested that if the troll responded with the first comment, the troll could remain in the community for the moment as “this person is at least trying to apologize.” But if the troll responded with the second comment, it would be more problematic and warrant a permanent banning of the troll, whereas the last comment would warrant a temporary banning. P11 noted, “it’s nice to plan some of these things out. I could even share this with other moderators.” 

\subsubsection{Role in the design process}
Many remarked that the generations were generally realistic (n=14). P26 noted, “it all felt really realistic… like the way they each spoke felt real. I’m assuming that someone (humans) actually wrote these, right?” P18 highlighted that they were pleasantly surprised by the details that were imbued in the generations after witnessing the generated members naming famous sites to visit in Pittsburgh in a community for “sharing and discussing about fun events around Pittsburgh,” while P24 even wanted to join the generated community for “learning about and creating AI generated art”, sharing that “this seems like a nice community. I’d probably subscribe to this community.” 

But this did not mean that our participants looked past the cracks in the generations’ realism. They shrewdly noticed that certain aspects of the generated posts and conversations were not likely (n=15) as P12 did: “I think this first one generally looks good, but it's a bit unrealistic in that I don't feel like people would actually use a long paragraph to say something about other humans.” They understood that “any tool is going to necessarily have limitations. I don't know how much I would trust that this would actually predict things that are likely to happen when I change this or that rule, just because humans are so unpredictable.”

Despite this, all 16 participants echoed that SimReddit adds to how they think about and craft their social designs. P5 noted, “I definitely think [the generations] add value to how you design… I’m impressed that just with the rules and the topic like these, it’s generating exactly what I’d expect to see.” Even P13, who started off skeptical in what the generations had to offer, shared, “This is actually more helpful than I expected!” But this was not because they believed that the generations predicted what is going to happen. Rather, they highlighted the tool’s ability to ground their assumptions about the community. P24 noted, “Sensemaking is a real challenge as a moderator… So [the generations] would give a good point of view to help you make sense of the unstructured barrage of comments that are coming and potentially become aware of things that seem to be in violation of [the community’s rules].” \mirev{Finally, the participants remarked on how SimReddit could supplement the ongoing efforts to form a better community. Some noted that these generations could be used to spark discussions between moderators about what content should be allowed (P8, P5), and aid communication between the moderators and other community members by using the generated conversations as  examples of what violates the community norms and rules (P13, P5). }

\subsubsection{Social simulacra for marginalized groups} \mirev{An important theme that arose in our designer evaluation was the social simulacra's role in designing for the marginalized groups in the community. Our evaluation included designers who identified as a member of marginalized groups during the study, including women of color (n=5) as well as religious and ethnic minorities (n=3) who experienced---and discussed in our interviews---misogyny, racism, and religious discrimination online. For both the designers from and outside marginalized groups, they used social simulacra to help them identify and describe the types of minority-targeting harassment that could arise in their designs. For instance, P9, a member of an ethnic minority designing a space for discussing non-fiction books, recognized from the simulacra community that one could send hateful messages against non-English speaking members by sharing literature with white supremacist themes. Meanwhile, P25, a male participant designing a space for urban exploration, learned from the simulacra community that comments taunting those weary of visiting dangerous places could readily turn misogynistic. Observations such as these encouraged our participants to add rules explicitly geared towards protecting these marginalized groups. In this context, GPT-3’s ability to generate harmful or bias content worked in our designers' favor as it could surface a large variety of such behavior that might target the designed community, beyond what the designers originally thought to protect against.

}
 
\section{Discussion}
In this section, we reflect on our contributions, ethical considerations, limitations, and opportunities for future work. 

\subsection{Designing With Social Simulacra}
As defined in the paper, the aim of social simulacra is to translate an idea for social design into a concrete set of plausible social interactions that might populate the design. There is enough realism in them to make them compelling and we can easily imagine that they might happen in the future. But social simulacra make no promises of perfect prescience. So why did our designers find them to be helpful aids as they envisioned and revised their social designs? 

Social simulacra fulfill the role that many of the early prototyping techniques fulfill outside of social computing design: they push the designer to question their assumptions, and through this, inspire them to consider a broader design space. Like any prototyping tool, social simulacra must be coupled with designer expertise: designers must discern which behaviors are probable and important enough for them to proactively act on. Our designer evaluation suggests that simulacra cue recall, with generated behaviors triggering recollections of what designers had seen in real life. \mirev{Compared to checklists, simulacra may cue this recall more effectively. Additionally, we believe that social simulacra can augment the existing practices such as participatory design by providing grounded examples of potentially problematic behaviors that require nuanced description (e.g., ``racial epithet'' is easy to explain in a checklist and imagine a design response, but ``context-specific micro-aggression'' is not) to facilitate more effective discussions amongst the designers and members of the community.} 

One implication of this work is a turn towards proactive, instead of reactive, design for social computing systems. Many of our interviewees commented on the negatives of only fixing mistakes once they already caused harm, and wished for opportunities to evaluate their designs ahead of deployment. They noticed flaws in the design largely by recognizing instances of harm to users, and then backtracking to find the underlying cause or coming up with a missing mitigation strategy. Such reactive strategies thus require harm to users and communities to improve; prototyping tools like ours can minimize the need for these strategies.



\subsubsection{False negatives} 
Social simulacra present additional trade-offs. Even with a multiverse, social simulacra are unlikely to present an exhaustive list of possible outcomes in a social system. Given this, the insights our designers get from social simulacra could have blind spots that are salient but not generated and thus not observed. This is analogous to the implied truth effect~\cite{pennycook2020implied}, where a tool to help identify possible issues falsely increases confidence that there are not other issues. On the other hand, owing to the breadth encoded in the training data of the large language model, prototyping via social simulacra is likely to produce more breadth than what any small collectives of test users would be able to provide.

\subsection{Limitations and Future Work}
As with any prototyping techniques, there are several limitations as well as future directions that are worth pointing out:

\subsubsection{Social simulacra will not predict the future}
Social dynamics of socio-technical system are complex and unpredictable~\cite{salganik2006exp}. And owing to this, regardless of how powerful the underlying model becomes, social simulacra will never have the ability to make a single point prediction on how a social system will develop. This might be viewed as a fundamental limitation of this approach, but we argue that this opens up a design space: our work presents one vision in the form of presenting many outcomes. But this design space is wide open for future interpretations. 

\subsubsection{Generalizing social simulacra beyond SimReddit}
\mirev{
Social simulacra is a general technique for leveraging generative models to prototype new social spaces. We focused on Reddit as an example, but we can use the technique to prototype any similar space (e.g., Facebook groups). Our implementation does not yet support some common features on social networks such as retweets, upvotes, likes, or shares; however, we can produce simulacra of many of these (e.g., likes, shares, etc.) using GPT-3 given its ability to classify (e.g., “will a Federer fan *like* a post about Federer winning Wimbledon?”). This allows us to also prototype algorithms that rely on engagement metrics. Finally, we expect the generalizability of the approach to expand as newer multi-modal models become available (e.g., DALL-E~\cite{RameshDalle} to prototype Instagram-like spaces). }

\subsubsection{Technical limitations}
There are a number of technical limitations represented in our implementation of social simulacra, SimReddit. Many of these will likely be surmountable. For instance, GPT-3 currently only accepts input prompts that are shorter than roughly 8,000 characters~\cite{Jagtap21}. In our implementation, this restriction occasionally means we must truncate earlier parts of a conversation, and that we could not feed the model a broader set of posts to convey overall norms in the space. Furthermore, SimReddit is currently text-only; however, in the future, if multi-modal generative models become more powerful, we can imagine opening up the types of content generated to include even multimodal posts (e.g., text and video) that are common on most social platforms today. Finally, SimReddit is limited to English language---and thus specific cultural contexts---because GPT-3 mostly operates in English. Future work could actively try to leverage new models to help designers bridge cultural gaps.

\subsection{Ethical and societal Impact}
\mirev{Beyond the limitations covered above, there are important ethical considerations to cover in regards to social simulacra. At an individual level, the generated content can be biased and problematic}~\cite{gehman2020toxic}. In some sense, within our prototyping use-case, this behavior is an asset to our tool; by replicating the bigoted and hateful behavior that has occurred online in the past, we allow designers to anticipate this behavior in their own contexts so they can design to account for these problems. However, this also means that social simulacra will expose designers to upsetting or triggering content. This is a delicate trade-off; in practice, communities often choose to accept this as the alternative may expose harmful content to both the designers and the community. 

At a societal level, there is a risk that this work may inspire tools for astroturfing, large-scale harassment attacks, and propaganda into the hands of malicious actors~\cite{38_bommasani, zellers2019defending}. \mirev{Given this, we advocate the following principles for releasing social simulacra. First, social simulacra should be usable only by vetted social computing designers. Second, one should centrally host and log the content that simulacra generate, so that their output can be audited to understand whether harmful attacks are being generated via the tool. Particularly, these systems should regularly sample from their generations and perform web searches to flag users who may be exporting this content en masse outside of prototyping scenarios.}

Social simulacra are also likely to replicate biases in who participates in online social spaces. Women and minoritized groups, for example, are often silenced in online spaces, and models such as GPT-3 may have learned these patterns and replicate them. This is problematic if it causes designers to then overlook these peoples' needs in design. In our prototypes, we mitigated this risk by ensuring a broad range of seed personas; this could become a more explicit test in future versions of our system. Social simulacra cannot and should not replace actual participatory engagement with users. There is a risk \mirev{that designers may overrely on simulacra}, which can lead to issues downstream. To mitigate this risk, prototyping via simulacra should be introduced in the broader context of human-centered design, which centers direct engagement.
\section{Conclusion}
In this paper, we present social simulacra, an approach enabling social computing designers to envision social behaviors that might occur in their space when populated. In this spirit, we created a SimReddit community dedicated to reviewing UIST papers. Here is what our generated reviewers had to say about our paper when given its abstract. 
Unfortunately R2, "a cynical reviewer inclined to reject the paper," reported: 
\begin{quote}
\gentxt{The idea of social simulacra is interesting, but the paper is poorly written and does not provide enough concrete detail about the technique.}
\end{quote}
But R1, "a social computing researcher who is thrilled about the paper," had a different view and responded: 
\begin{quote}
\gentxt{This is an excellent paper. The authors have developed a technique that will be of great use to designers of social computing systems. The paper is well-written and the technique is clearly described and illustrated with examples. I believe this paper makes a significant contribution to the field of social computing.}
\end{quote}
And last but not least, AC, "a senior faculty member who didn’t have time to write a full review," noted: 
\begin{quote}
\gentxt{This is a thought-provoking paper. The idea of social simulacra is intriguing and the paper provides a detailed description of the technique. However, I would have liked to see more discussion of the potential benefits and limitations of the approach.}
\end{quote}
We sincerely thank the (generated) reviewers for their invaluable time, comments and suggestions.









\begin{acks}
We thank Michael Terry and Mitchell L. Gordon for their insights and comments on drafts of this paper. We also thank the Stanford Human-Centered AI Institute (HAI), Google Research, the Hasso Plattner Design Thinking Research Program (HPDTRP), and \mbox{OpenAI} for their funding support. 
\end{acks}

\bibliographystyle{ACM-Reference-Format}
\bibliography{main}

\end{document}